\documentclass[]{CHRONOS_s}  
\usepackage[]{graphicx}
\usepackage[]{url}

\title{Development of epoxy-based millimeter absorber with expanded polystyrenes and carbon black} 

\author{Yuki Inoue\supit{a,b,c,d},Masaya Hasegawa\supit{d,e,f},Masashi Hazumi\supit{d,e,f,g,h},Suguru Takada\supit{i},Takayuki Tomaru\supit{e,f,j}
\skiplinehalf
\supit{a}Department of Physics, National Central University, Zhongli, Taoyuan 32001, Taiwan; \\
\supit{b}Center for High Energy and High Field Physics, National Central University, Zhongli, Taoyuan 32001,Taiwan; \\
\supit{c}Institute of Physics,  Academia Sinica, Taipei 11529, Taiwan; \\
\supit{d}High Energy Accelerator Research Organization, Ibaraki 305-0801, Japan; \\
\supit{e}SOKENDAI (The Graduate University for Advanced Studies), Ibaraki 305-0801, Japan.; \\
\supit{f}International Center for Quantum-field Measurement System for Studies of the Universe and Particle(QUP), High Energy Accelerator Research Organization(KEK), Tsukuba, Ibaraki 305-0801, Japan.; \\
\supit{g}Kavli Institute for the Physics and Mathematics of The Universe (WPI), The University of Tokyo, Chiba, 277-8583 Japan.; \\
\supit{h}Institute of Space and Astronautical Science (ISAS), Japan Aerospace Exploration Agency (JAXA) Kanagawa 252-5210, Japan.; \\
\supit{i}National Institute of Fusion Science, Gifu, 509-5202, Japan; \\
\supit{j}National Astronomical Observatory of Japan, Tokyo, 181-8588, Japan.; 
}

\authorinfo{Further author information: Yuki.Inoue.: E-mail: iyuki@ncu.edu.tw, Telephone: +886-3-422-7151}

\begin{document} 
  \maketitle 

\begin{abstract}
We recently developed and characterized an absorber for millimeter wavelengths. To absorb the millimeter wave efficiently, we need to develop the low reflection and high absorption material. To meet these requirements, we tried to add polystyrene beads in the epoxy for multi-scattering in the absorber.
The typical diameter of polystyrene beads corresponds to the scale of Mie scattering for the multi-scattering of photons in the absorber. The absorber consists of epoxy, carbon black, and expanded polystyrene beads. The typical size of the expanded polystyrene beads is consistent with the peak of cross-section of Mie scattering to increase the mean free path in the absorber. By applying this effect, we succeeded in improving the performance of the absorber. In this paper, we measured the optical property of epoxy for the calculation of the Mie scattering effect. Based on the calculation result, we developed the 8 types samples by changing the ratio of absorber material. To compare 8 samples, we characterized the reflectance 
and transmittance of the absorber in millimeter length. The measured reflectance and transmittance of 2 mm thickness sample with optimized parameter are less than 20\% and 10\%. We also measured the transmittance in sub-millimeter wavelength. The measured transmittance is less than 1\%. The shape of absorber can be modified for any shape, such as chip and pyramidal shapes. By using this absorber, we can apply for the mitigation of stray light of millimeter wave telescope with any shapes. 
\end{abstract}


\keywords{Absorber, Infrared, Blackbody, Cosmic Microwave Background, Millimeter wave, Gravitational Waves , POLARBEAR-2, LiteBIRD}

\section{Introduction}
\label{sec1}


Progress in the development of millimeter wave technologies has been successful in various fields, such as astrophysics, defense, security identification, stealth technology, calibration and several more~\cite{review,ed,iso}. The potential and possibilities of millimeter waves are also expected to improve future technologies. Therefore, the development of a high performance millimeter absorber is essential. In particular, an absorber with a high absorption and low surface reflectance can be applied to attenuate the unexpected stray light and aperture of millimeter beams.
Furthermore, high absorbing material can be applied to a black body radiation source. Many experiments have employed the high-performance black body to calibrate and characterize the experimental systems~\cite{review}.

One of the most interesting applications of a millimeter absorber is for the cosmic microwave background (CMB) polarization experiment~\cite{1, 2}. CMB experiments often use the cryogenic millimeter absorber to reduce stray light~\cite{3,4,5}, as well as for the design of beam apertures. The performance of an absorber as a black body can be applied for calibrating cosmological experiments~\cite{6, 7}.

One of the typical material structures of a millimeter absorber is the open-cell foam, such as HR-10 or AN 72 made by Emerson \& Cuming~\cite{8}.
closed-cell absorbers are also commonly used, such as bock black~\cite{9} or CR-112~\cite{8}. The open-cell foam absorber features the advantages of low reflection and high absorption. However, the typical thermal conductivity of the open-cell foam absorber is not sufficient for application in a cryogenic environment~\cite{6,cry}. This is because the open-cell structure effectively reduces thermal conductance. Moreover, open-cell foam machinability cannot make any shape of aperture. As a result, CMB experiments prefer to use the closed-cell absorber because of sufficient cooling with thermal conduction~\cite{cry}. We can easily design the shape of a closed-cell material absorber. However, these types have high reflectance in the millimeter wavelength due to their high index of refraction (IoR). To prevent high reflectance on the surface of a millimeter absorber, sometimes a rough surface is created to absorb the millimeter wave. However, these types have high reflectance in the millimeter wavelength due to high index of refraction. To increase the absorption of material for black body application, multi-reflection is often employed with pyramidal, conical, and taper shapes~\cite{conical,ed,fixen}. Therefore, to improve the property of an absorber, we require high absorption, low reflectance, high machinability, and high thermal conductivity.

In this paper, we introduce an improved absorber. This absorber has high absorption, low reflectance, and is easy to fix to a variety of forms. To obtain these properties, we used expanded polystyrene, so called powder beads, with epoxy and carbon black. The combination of epoxy and carbon black has been used in previous studies. According to previous research, the absorption of carbon at millimeter wavelength meet the requirement of absorption in this application~\cite{carbon}.   By mixing powder beads, we succeeded in adding new properties. The diameter of the powder beads is controlled to the Mie-scattering scale of millimeter wavelength. By applying the effective medium theory, we can reduce the effective IoR of the surface on the absorber~\cite{surface, meta, 3D,comp}. In this paper, we explain a new type of absorber that is loaded with high absorption, low reflectance, and designable shapes. In Section~\ref{sec2}, we describe the material property and selection. In Section~\ref{sec3}, we explain the requirement of the absorber and characterization of the absorber. In Section~\ref{sec4}, we discuss the results of measurement and application.
\begin{figure}[b!]
\centering
\includegraphics[width=.77\linewidth]{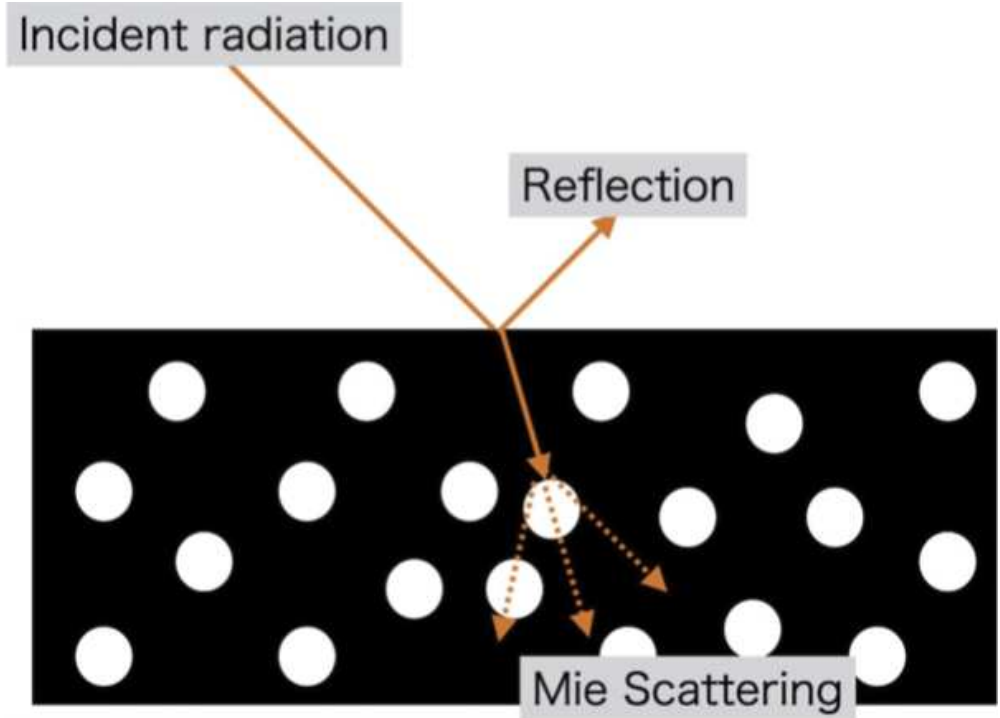}
\caption{Cross section of KEK black. Surface reflection is defined by the IoR of the base material. White circles correspond to powder beads as a scattering material. The scattered photon is absorbed by the carbon black as an absorber.The peak of cross-section of Mie scattering is corresponding to the typical size of powder beads. It can increase the mean free path length in the absorber. Numerical calculation of scattering effect is discussed in Sec.~\ref{Scat}. The scattered photon is absorbed by the carbon black.}
\label{fig1}
\end{figure}
\section{Design and fabrication}
\label{sec2}

KEK black\footnote{KEK black is named after the name of the developed institute, high energy accelerometer research organization (KEK). In this paper, we call it as KEK black. } consists of the base material, absorbing material, and scattering material. We employed low IoR epoxy and carbon as a base material and absorber. We then selected expanded polystyrene, which is a low IoR material in millimeter wave, as the scattering material. The scattering material is used to amplify absorption with multi-scattering. Previous absorbers have been made with a higher IoR material than that of the base material. However, we focused on the difference of IoR between the base material and the scattering material because the scattering rate is the only function of differences of the two materials. Therefore, we decided to use expanded polyethylene material as a scattering material.

Figure~\ref{fig1} is a schematic view of KEK black. POLARBEAR-2 and Simons Array experiments also employed KEK black for a millimeter absorber and beam aperture~\cite{3, 10}. The requirements of POLARBEAR-2 for an absorber were:


\begin{itemize}
\item loss tangent is higher than 0.3
\item index of reflection is less than 1.55.
\end{itemize}

This is because the black body is assumed to have higher than 99\%
absorption and lower than 5\% reflection when we mount the 5~mm-thick black body on the metal plate.

We provide details of each material in the following sub-sections.

\begin{figure}[ht!]
\centering
\includegraphics[width=\linewidth]{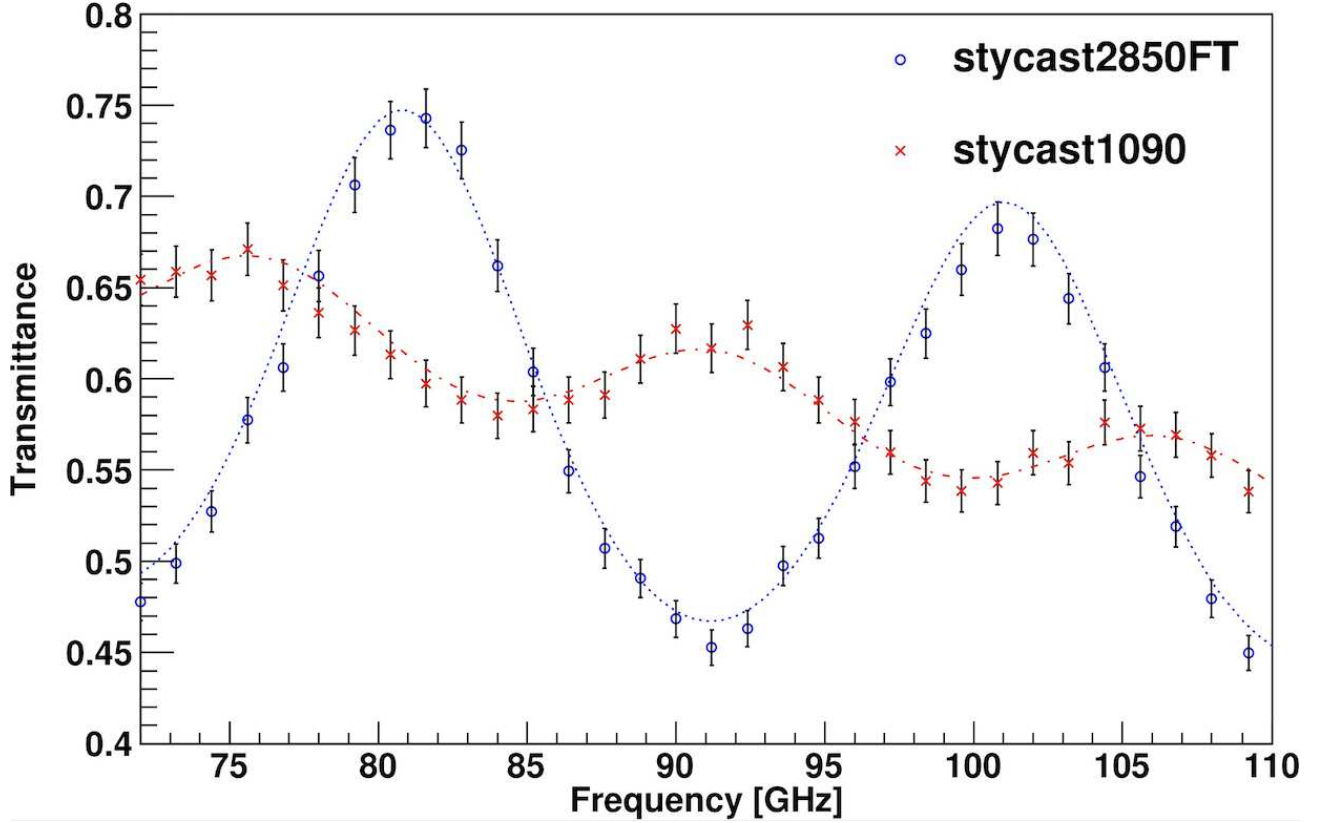}
\caption{Transmittances of the two types of epoxies as a function of frequency in the millimeter-wavelength range. Sample temperature was 300~K. The red and blue points are samples of Stycast 2850FT and Stycast1090, respectively. The dominant error is due to the gain drift of the amplifier in the synthesizer, which arises from the temperature fluctuation. The two curves are the best fits of each sample. The results of the fits are listed in Table~\ref{tab1}.}
\label{fig2}
\end{figure}

\subsection{Base material}

To optimize the design of KEK black, we characterized the material property of epoxy as a base material. In this paper, we focused on stycast2850FT and stycast1090 made by Emerson \& Cuming~\cite{8, 11,13}. Both materials are often used in cosmic microwave background experiments and astronomical telescopes. The thermal expansions of these epoxies are controlled to the thermal expansion of brass at low temperatures. These epoxies are also used for the anti-reflection coating material of millimeter wavelength. In this paper, we compared the properties of two materials. The reflection of the absorber is described as
\begin{equation}
R=r^2\frac{1-2\mu^2 \cos\phi + \mu^4}{1-2r^2\mu^2 \cos\phi + r^4\mu^4},
\label{eq1}
\end{equation}
where $R$ and $n$ are reflectance and IoR, respectively~\cite{14}. The transmittance of the bulk material is obtained as
\begin{equation}
T = \frac{t^4 \mu^2}{1-2r^2\mu^2 \cos \phi +r^4\mu^4},
\label{eq2}
\end{equation}
where
\begin{eqnarray}
t^2 & = \frac{4n}{(1+n)^2}, \\
r^2 & = \frac{(1-n)^2}{(1+n)^2}, \\
\mu & = {\rm e}^{-k n d \tan \delta }, \\
\phi & = k n d.
\label{eq3}
\end{eqnarray}

Transmittance $T $ is expressed in terms of IOR $n$, loss tangent tan $\delta $, thickness $d$, and wavenumber $k =  2 \nu \pi c $~\cite{14}. To characterize the performance of epoxy as an absorber material, we had to estimate the IoR and loss tangent. The measured spectra of the two epoxies between 72 and 108~GHz are shown in Figure~\ref{fig2}. We fit spectra to estimate IoR and loss tangent based on~ Eq.\ref{eq2} by fixing measured thickness $d $ and frequency $\nu $. The result of IOR and loss tangent measurements are shown in Table~\ref{tab1}.  The measured parameters are consistent with the previous measurement from literature~\cite{Toki}. In this literature, they measure the same epoxies with  Michelson Fourier transform spectrometer. According to these results, we decide to use Stycast 1090 for the epoxy material because of its high loss tangent and low IoR.

\begin{table}[h!t]
\centering
\caption{IoR and loss tangent of epoxy between 72 and 108~GHz for each 1.2~GHz step. We fit the transmission spectrum by using~\ref{eq2}. The dominant uncertainty is due to the gain drift of the amplifier in the synthesizer, which arises from the temperature fluctuation.}
\label{tab1}
\begin{tabular}{cccc}
\hline
Name&Thickness~[mm]&Index&Loss-tangent~[$\times 10^{-2}$]  \\
\hline
Stycast 2850FT  &$3.38\pm 0.01$&$2.194\pm 0.006$&$1.88\pm0.04$ \\
Stycast 1090  &$6.85 \pm 0.01$&$1.435 \pm 0.005$&$ 2.46\pm0.01$ \\ 
\hline
\end{tabular}
\end{table}
\begin{figure}[ht]
\centering
\includegraphics[width=\linewidth]{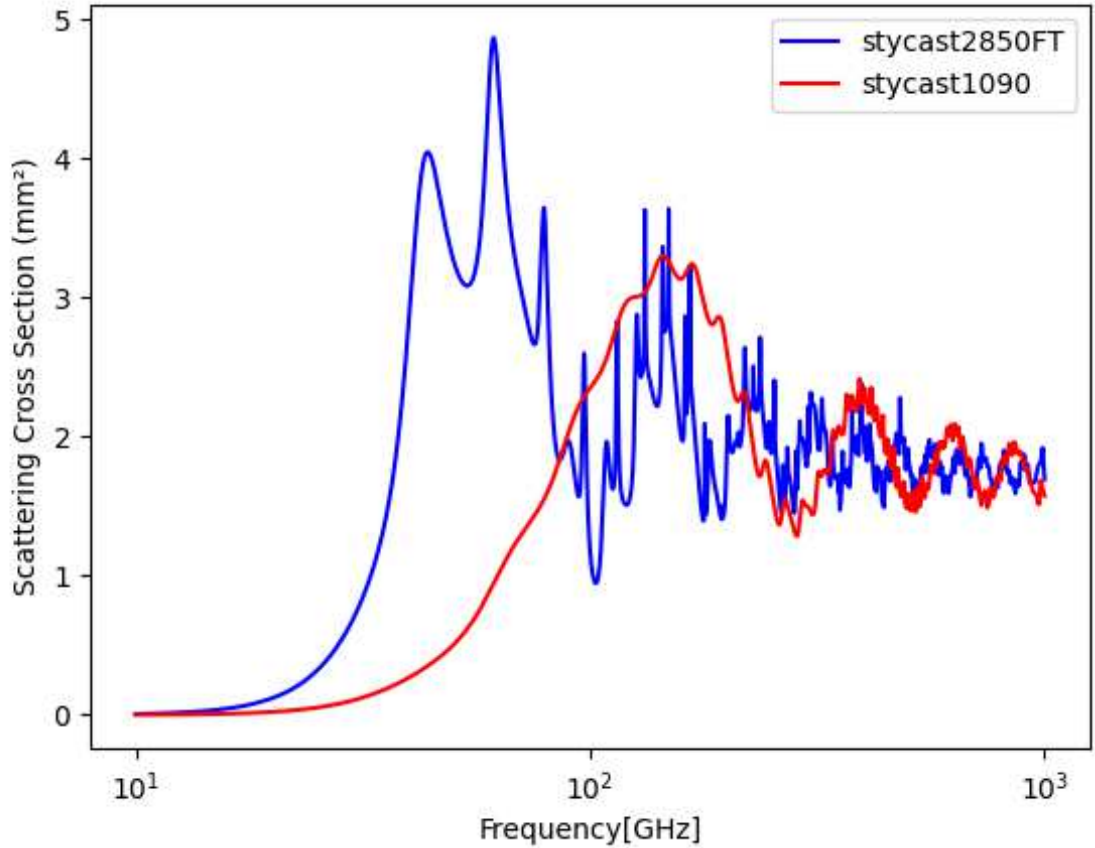}
\caption{Analytical calculation of Mie scattering. We calculate the cross-section by assuming the powder beads diameter and the measured material propertes of Stycast2850FT and Stycast1090.}
\label{mie}
\end{figure}

\begin{figure}[ht]
\centering
\includegraphics[width=\linewidth]{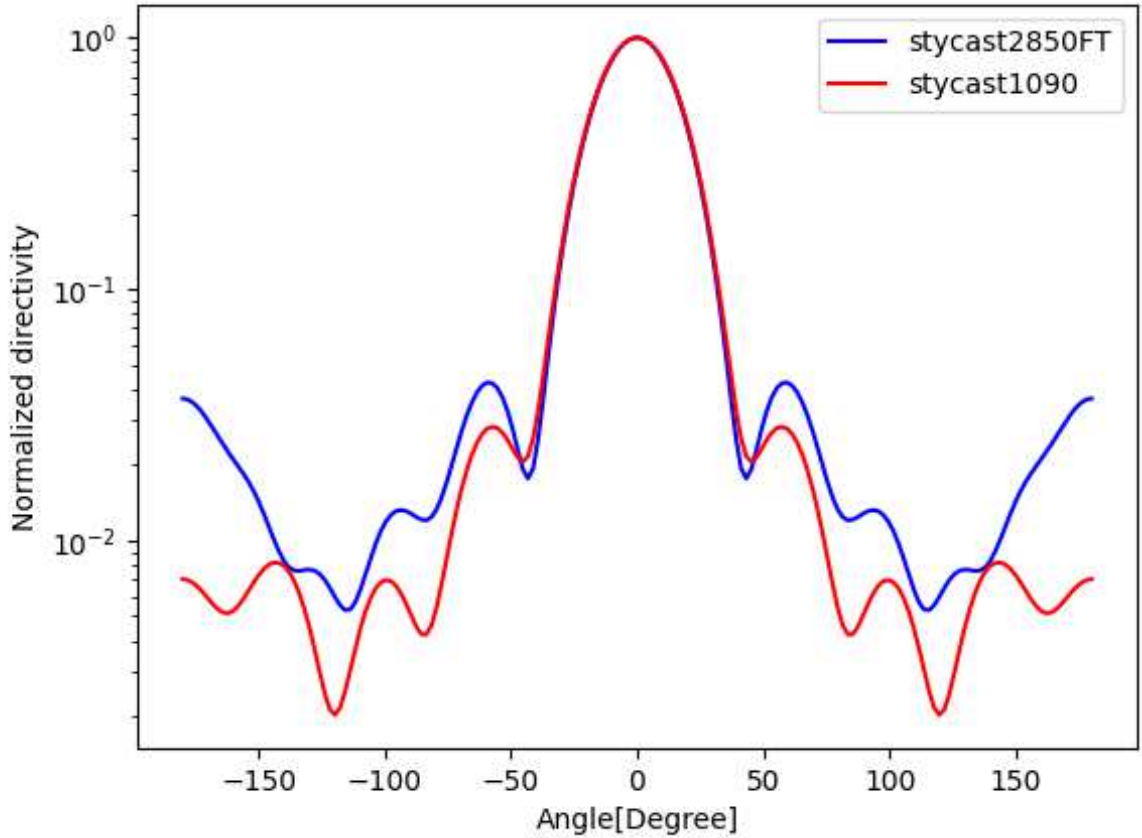}
\caption{Normalized directivity of Mie scattering. By assuming the parameter in Fig.~\ref{mie}, we calculate the directivity at 150GHz signal.}
\label{pole}
\end{figure}
\subsection{Absorber material}

According to a previous study by James Bock~\cite{9}, the researchers succeeded in increasing the absorption of epoxy by mixing in carbon lump black. In our study, we used carbon black for the absorbing material. The carbon black used was obtained from Mitusi Kagaku~\cite{15}. Both carbon black and lump black are fabricated using the oil-furnace method, but the difference between lump black and carbon black consists of the properties of oil. Lump black is made from burned vegetable oil, while carbon black is made from burned mineral oil. Ultimately, we used carbon black as an absorption material because the performances between both materials did not differ significantly.

\subsection{Scattering material} \label{Scat}

Powder beads are expanded polystyrene used for stuffing a doll, cushion, or pillows made by the MOGU company~\cite{16}. The typical diameter of a powder beads is 0.5~mm, which is near the typical size of Mie-scattering in the millimeter wavelength. When millimeter waves enter the absorber, they scatter several times. The multi-scattering effectively increases the optical path length in the absorber. The extended optical path length by Mie-scattering increases the effective absorption of the absorber. Figure~\ref{mie} and \ref{pole} shows the numerical calculation of Mie scattering for stycast2850FT and stycast1090 cases. The numerical calculation is given by  miepython~\cite{mie}. assumed the size of power beads and index of refraction of Stycast1090 an Stycast2850FT. The peaks of Mie scattering exist around 100 GHz region. We also calculate the normalized directivity at 150GHz for each case. To induce the effective scattering, the size of power beads are reasonable.

\subsection{Fabrication method}

\begin{figure}[ht!]
\centering
\includegraphics[width=.9\linewidth]{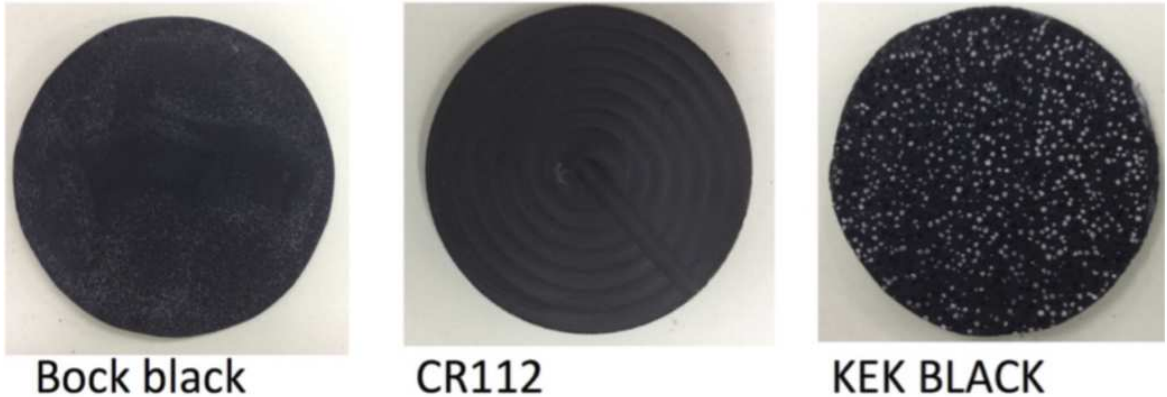}
\caption{Pictures of different absorbers. The figure on the right shows the improved absorber. The middle figure shows CR112. The figure on the left shows Bock black.}
\label{fig3}
\end{figure}

In this section, we explain our fabrication method for KEK black. In the first step, we mixed the Stycast 1090 and catalyst 9 for a couple of minutes. Next, carbon black powder was mixed together with the epoxy, and we kneaded it until it was smooth like clay. In the third step, we mixed the powder beads into the clay-like epoxy with carbon black. Finally, we put the clay-like absorber in the mold and waited 24~hours until it was cured. The recipe of the measured sample is listed in Table~\ref{tab2}. We made eight types of KEK black with different compositions of the carbon black and powder beads. Furthermore, we made samples of the Bock black~\cite{9} and CR112~\cite{8} to compare with typical absorbers in CMB experiments as shown in Fig.~\ref{fig3}. Bock black is developed for the application of absorber for near-infrared region. It includes the microsphere for the increasing multi-scattering for absorption. However, material of sphere is silica and typical size is 1000 times less than that of KEK black. CR112 is also for the absorber of millimeter wavelength. It contains the powder of steel for the effective absorption.

\begin{table}[h!t]
\centering
\caption{Recipes of the measured samples. We made eight samples by changing the carbon black and powder beads ratio.}
\label{tab2}
\begin{tabular}{ccccc}
\hline
Sample name & Stycast1090 [g] & Catalyst9 [g] & Carbon black [g]   & Powder beads [g] \\
\hline 
A      & 10          & 0.9       & 2         & 0.0 \\
B      & 10          & 0.9       & 2         & 0.1 \\
C      & 10          & 0.9       & 2         & 0.5 \\
D      & 10          & 0.9       & 2         & 0.7 \\
E      & 10          & 0.9       & 4         & 0.0 \\
F      & 10          & 0.9       & 4         & 0.1 \\
G      & 10          & 0.9       & 4         & 0.5 \\
H      & 10          & 0.9       & 4         & 0.7 \\
\hline
\end{tabular}
\end{table}

\section{CHARACTERIZATION OF ABSORBER}
\label{sec3}

The performance of the absorber is characterized by the measurements of transmittance and reflectance. We explain measured transmittance and reflectance based on the measurement system with Fourier-transform spectrometer and frequency multiplier with millimeter source. To compare the performance of each parameter as listed in Table~\ref{tab2}, we made samples with the same shapes.

\subsection{Transmittance in millimeter wavelength}

\begin{figure}[h!t]
\centering
\includegraphics[width=.9\linewidth]{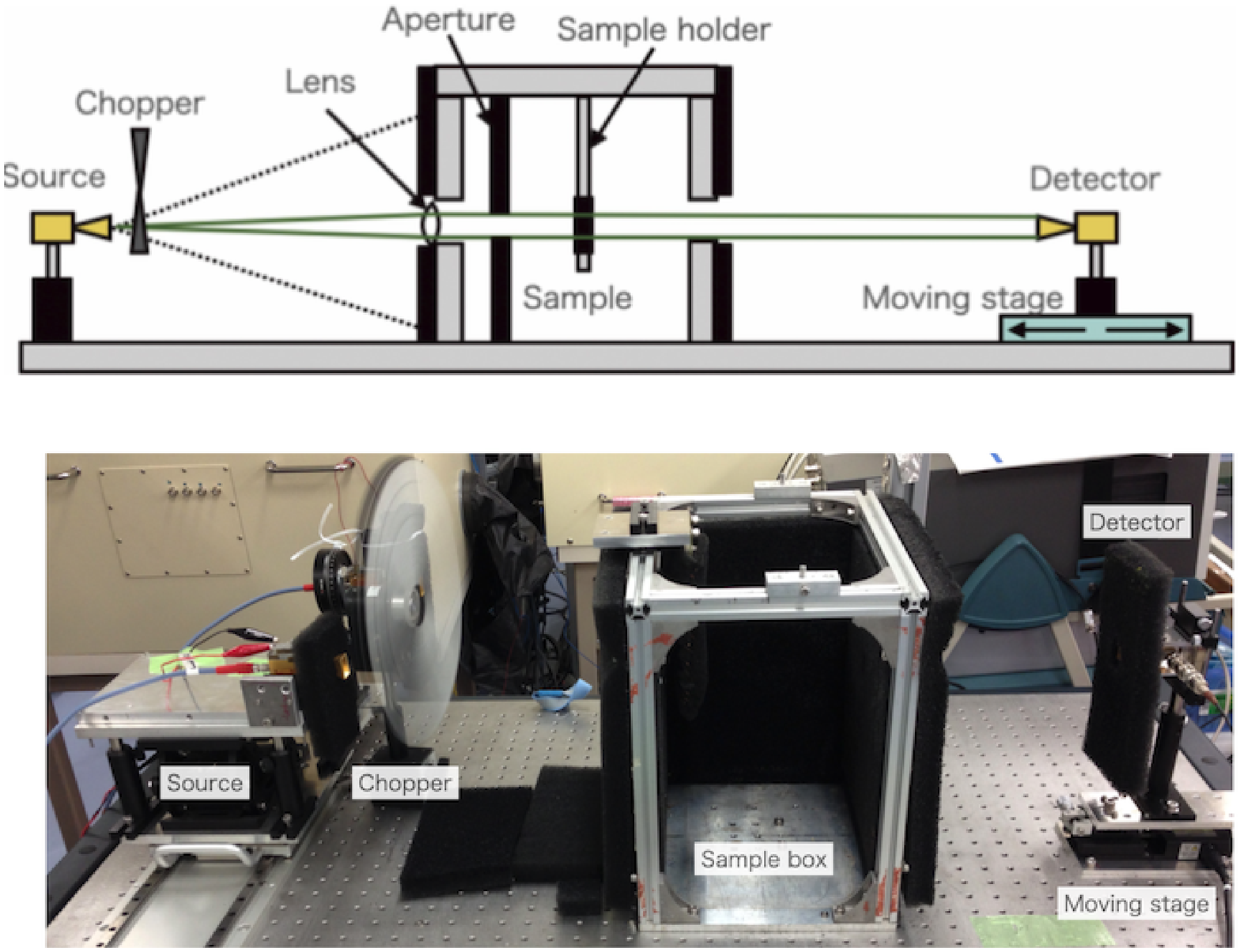}
\caption{Schematic view(top) and picture(bottom) of transmittance measurement system. The millimeter wave is emitted from the source. We chopped the beam to attenuate the effect of the time drift. The beam is collimated with the lens and aperture. We measured the transmittance by mounting the sample on the holder.}
\label{fig4}
\end{figure}

To measure the performance of the absorber, we made sample discs that were 2~mm in thickness and 50~mm in diameter. The measured frequencies were between 72 and 146~GHz. To cover the frequency range, we employed $\times $9 and $\times $6 frequency multipliers with a signal generator~\cite{17, 18}. The setup of the measurement system is shown in Figure~\ref{fig4}. The signal is chopped at 15 Hz for modulation with a lock-in amplifier, and the modulated signal is obtained using a DAQ. Detail of measurement system is described in elsewhere~\cite{11,17}.

\begin{figure}[h!t]
\centering
\includegraphics[width=\linewidth]{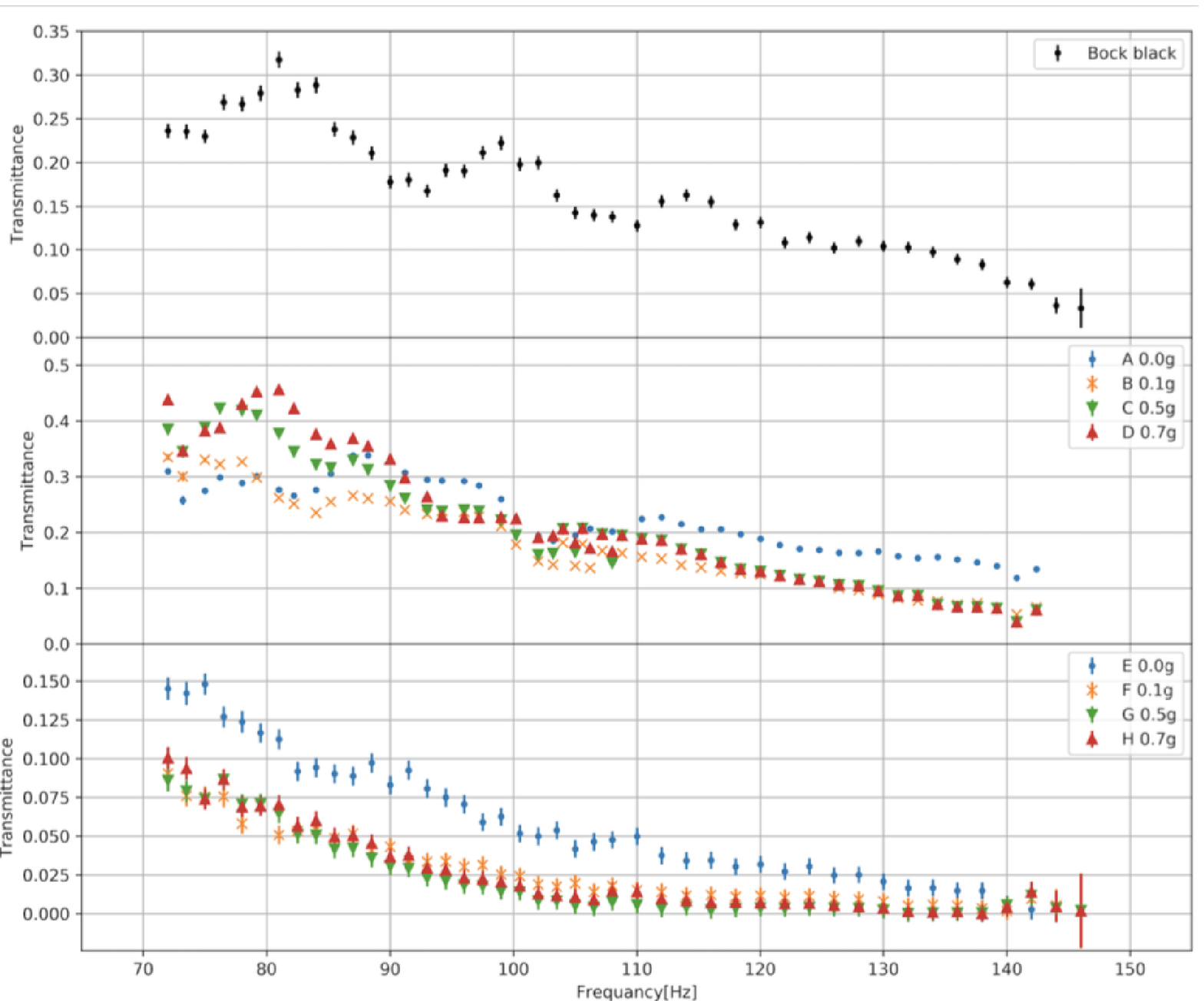}
\caption{The measured transmittance of KEK black. The measured frequencies are between 72 and 146~GHz. Each error is statistical due to gain fluctuations. The first panel shows the transmittance of Bock black, where the error bars show $\pm 1\sigma $ uncertainty. The second to third panels show the results of KEK black.}
\label{fig5}
\end{figure}

We mounted the horn on the optical post and illuminated the sample through the aperture. To define the aperture hole, we used HR10, which was the open-cell foam black body. We also used an ultra-high density polyethylene lens for the collimation of the beam. The beam was chopped by the optical chopper to reduce the noise and drift of the gain during measurement. When we processed the data, we measured the fringe pattern of the samples by changing the detector position, as shown in Figure~\ref{fig4}. This process enabled us to mitigate the multi-reflection effect in the sample. The measured transmittance is shown in Figure~\ref{fig5}.

The transmittance values of samples A, B, C, and D are higher than bock black because of the small amount of carbon. However, when we increase the carbon black from 2~g to 4~g, the transmittance values of samples E, F, G, and H are less than bock black. Therefore, the performance of E, F, G, and H was better than A, B, C, and D. The transmittance of all samples with powder beads was decreased. Furthermore, the transmittance of group E, F, G, and H was less than that of group A, B, C, and D because of the cross section of light and carbon, suggesting that Mie-scattering effectively increases the optical path length of light.
However, we observed no significant differences between C and D (G and H). When we ground the surface of KEK black for H, it occasionally cracked. As a result, G has the advantage of machinability of these four samples.

\subsection{Reflectance measurement}

\begin{figure}[h!]
\centering
\includegraphics[width=.85\linewidth]{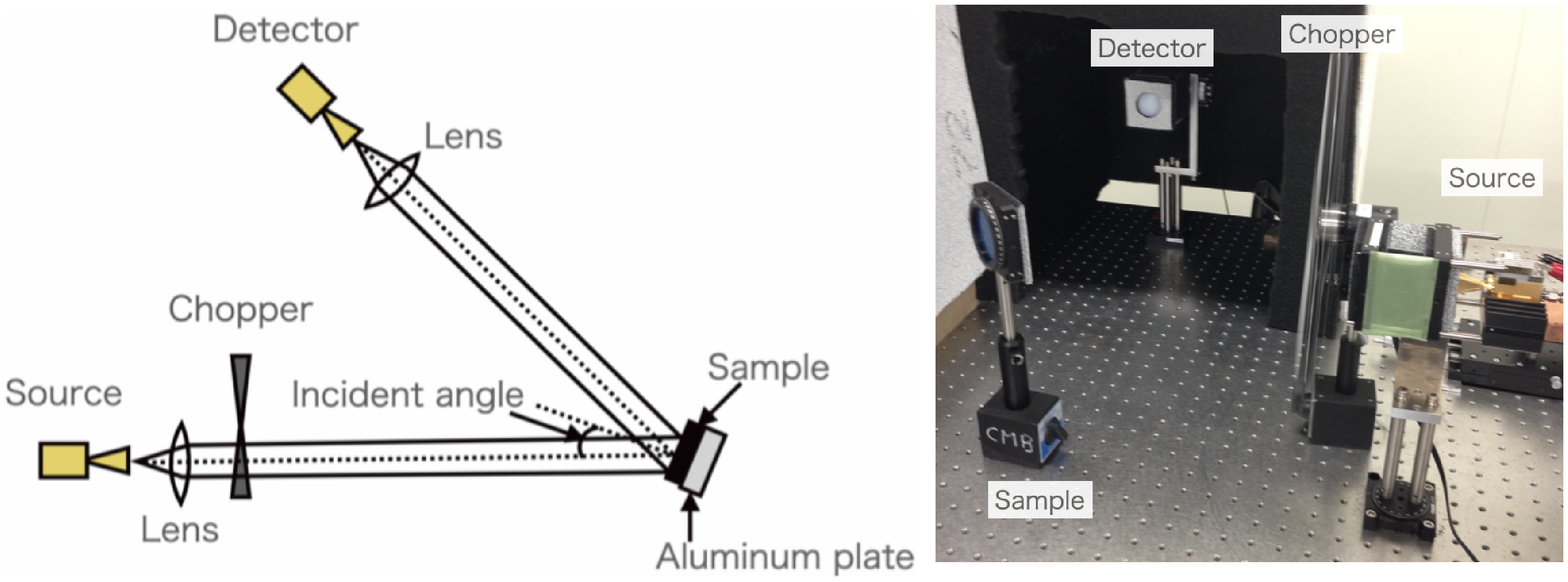}
\caption{Schematic view(left) and picture(right) of the reflectance measurement system. The source and detector are mounted on the optical table at an angle of 60~degrees. The beam is collimated with the UHMWPE lenses and is reflected by the surface of the measured sample, which is placed on the aluminum plate. The signal is modulated by a chopper whose frequency is 13~Hz.}
\label{fig6}
\end{figure}

\begin{figure}[h!]
\centering
\includegraphics[width=\linewidth]{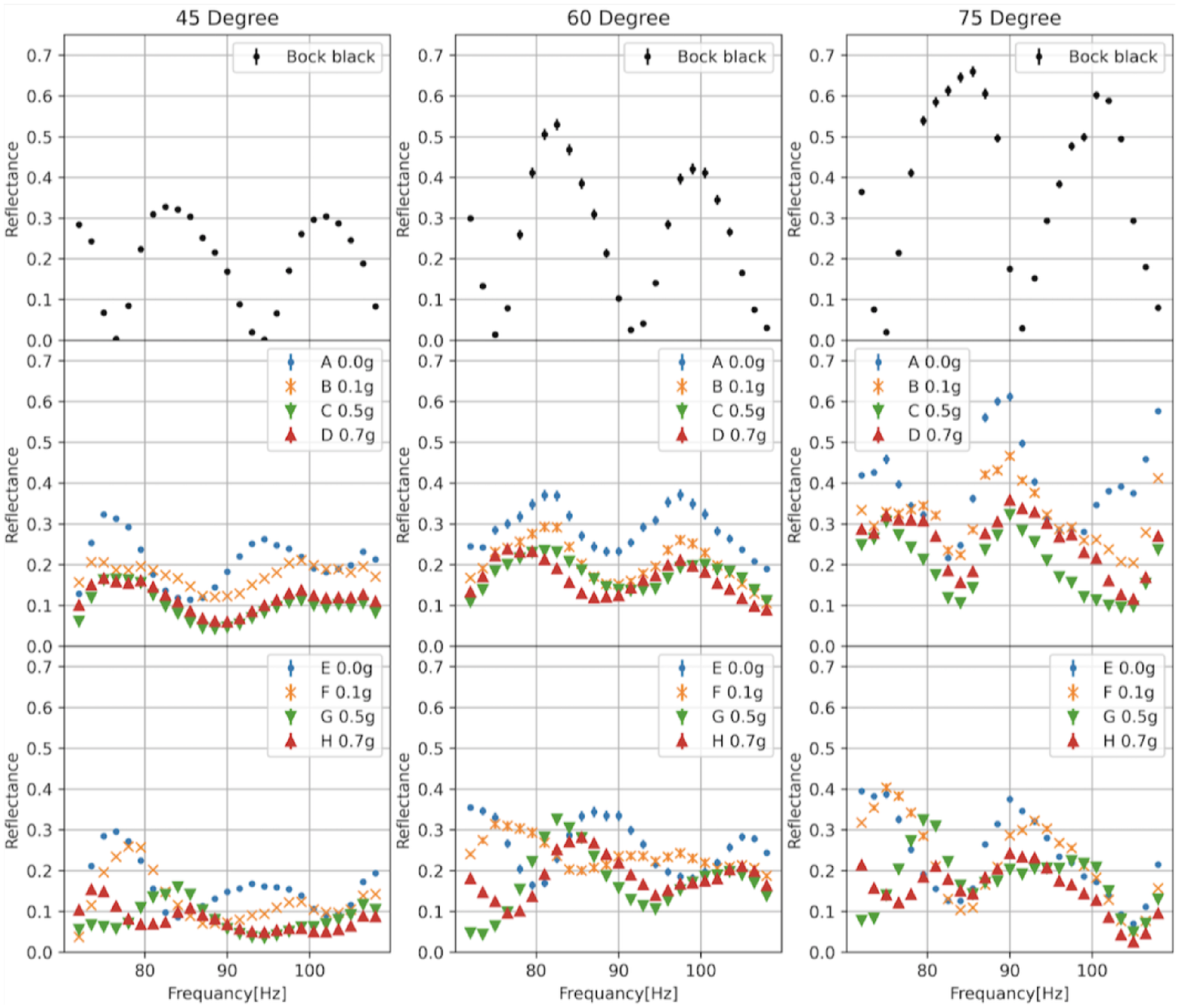}
\caption{Reflectance with 45~degrees(left), 60~degrees(middle), and 75~degrees(right) for the Bock black and KEK blacks as a function of frequency in the millimeter-wavelength range. Sample temperature was 300~K. The samples are listed in Table~\ref{tab2}. The dominant error was due to the gain drift of the amplifier in the synthesizer, which arises from temperature fluctuations.}
\label{fig7}
\end{figure}

Because of the conservation law of energy, characterization of bulk reflectance is an important parameter for absorbers. We can usually mount the absorber on a metal surface and neglect transmittance. Then, absorption of bulk, $A$, is obtained as $A =  1 - R$, where $R $ is the bulk reflectance. Therefore, low reflectance absorbers are preferred. In this section, we developed the reflectance measurement system of a sample on a metal plate. The diameter and thickness of the samples were 50~mm and 2~mm, respectively. We mounted them on the optical mount for each incident angle. The measurement system is shown in Figure~\ref{fig6}. We employed the sixfold frequency multiplier with the signal generator, and the frequency range of the signal generator was between 12 and 18~GHz. Thus, the multiplied frequency was between 72 and 108~GHz. We measured the spectrum with 0.12~GHz frequency resolution. The beam was collimated with lenses and HR-10 apertures, and the reflected beam was detected by a diode detector. The signal was chopped at 13~Hz for heterodyne detection at the modulated frequency with a lock-in amplifier, and the signal was recorded using a data logger. The detector was actuated along the optical axis for more than a half wavelength of fringes to subtract the effect of a standing wave in the measurement setup. The measured reflectance is shown in Figure~\ref{fig7}.

Fringes of the measured data were caused by the multi-reflection between the surface of the absorber and the metal plate. In general, fringe with multi-reflection in the material was attenuated by the loss in a material.
We can attenuate the fringe effect by increasing the thickness of the sample. However, the mean of reflection is not mitigated because the surface reflection is not a function of thickness and loss. Therefore, surface reflection can be one of the parameters to explain the effectiveness of an absorber. According to Fig.~\ref{fig7}, the surface reflection of bock black at the millimeter wavelength is higher than that of KEK black. Reflectance values at 45$^{\circ}$, 60$^{\circ}$, and 75$^{\circ}$ correspond to 15\%, 30\%, and 35\%, respectively. This finding is because the IoR of KEK black is less than that of bock black. We used Staycast1090 for the epoxy in the absorber. However, Bock black used Stycast2850FT. The expected surface reflections of the two epoxies were calculated using~\ref{eq1} and are shown in Table~\ref{tab1}. The calculated surface reflectance values of Stycast2850FT and Stycast1090 were 14.0\% and 3.2\%, respectively, which explains why the surface reflection of KEK black is less than that of Bock black. The powder beads also support the attenuation of multi-reflection because of the mean free path length. In particular, the surface reflection of samples C and D (G and H) was reduced.
This result is consistent with the transmittance measurement. According to the above discussion, we conclude that C or G has the best advantage as a millimeter absorber.

\subsection{Transmittance at the sub-millimeter wavelength}

\begin{figure}[ht!]
\centering
\includegraphics[width=\linewidth]{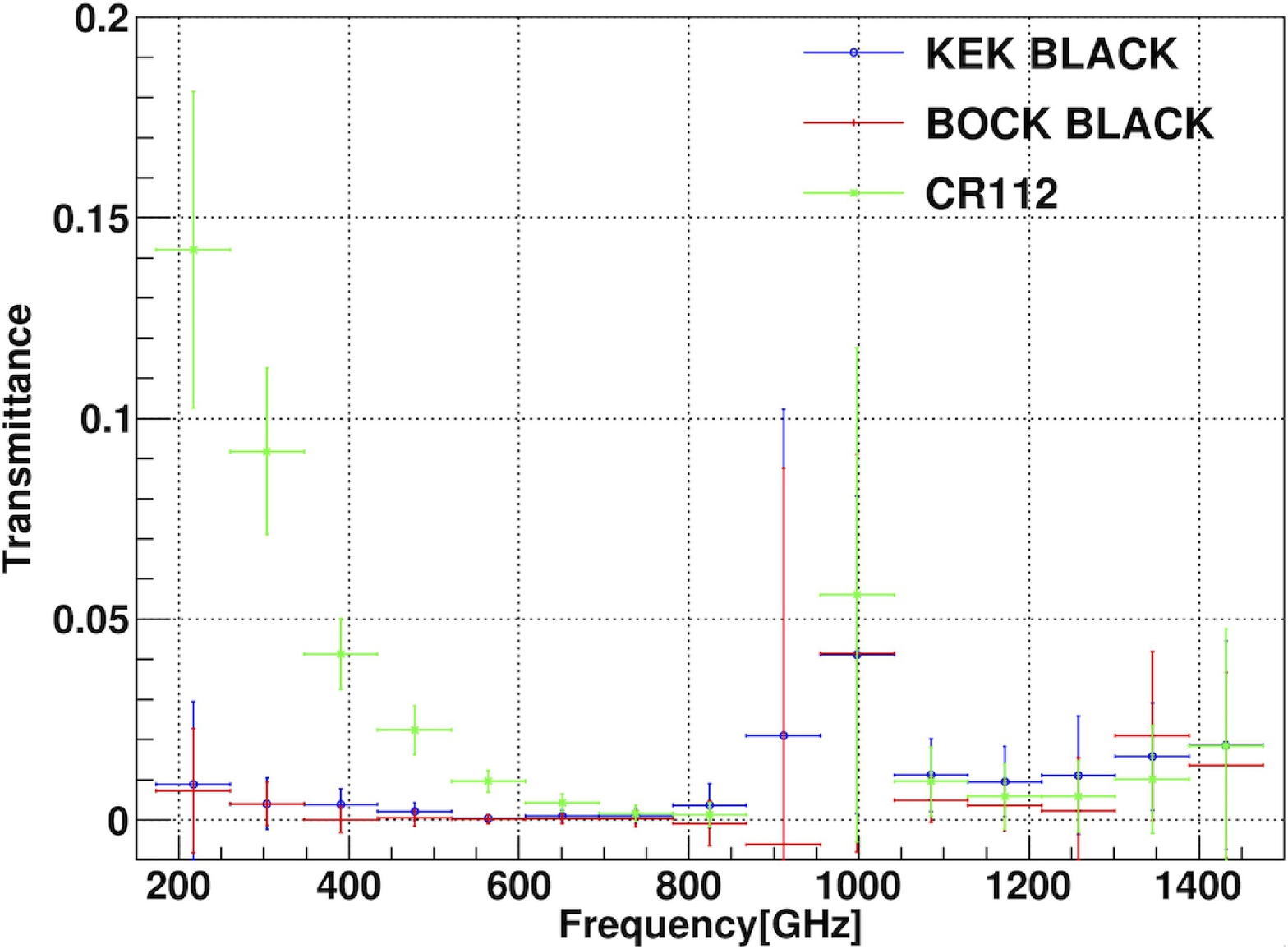}
\caption{the measured transmittance of the absorber. The measured samples are KEK black, Bock black, and CR112. The sample thickness and diameter are 2~mm and 20~mm. Measurements were carried out at 300~K.}
\label{fig10}
\end{figure}

Application in a sub-millimeter telescope is another possibility for KEK black. According to the above measurements, we concluded that the performance of sample G is better than that of the others. We compared sample G with such popular absorbers as CR110 and Bock black in the sub-millimeter wavelength. The measured samples were KEK black, bock black, and CR112, as shown in Figure~\ref{fig3}. We measured the transmittance from 250~GHz to 1500~GHz at 300~K with Fourier-transform spectrometer system~\cite{19}. The transmittance spectra of 2~mm thick samples is shown in Figure~\ref{fig10}. We mounted the 2~mm sample on the sample holder with a detector. By injecting the source of the mercury lamp with a Michelson type interferometer, we were able to detect the interferometric signal by changing mirror displacement. The transmittance spectra of the 2~mm thick samples are shown in Figure~\ref{fig10}.

We observed no significant differences between KEK black and Bock black. However, the transmittance of CR112 at the low frequency was larger than that of others. The wavelength of the sub-millimeter range is smaller than the diameter of the powder beads. However, the bubble structure of the powder beads caused the multi-scattering in the absorber. This effect increases the effective mean free path and absorbed it. Still, the directivity of Mie-scattering is greater than Multi-scattering. Therefore, we should control the typical diameter of the bubble to be Mie-scattering scale for the target wavelength. On the other hand, Bock black uses a microsphere whose diameter is optimized for sub-millimeter and Infrared wavelengths. Therefore, Mie-scattering in Bock black can effectively absorb sub-millimeter wave light.

%
\section{Discussion}
\label{sec4}

The absorber in this study included a base material, an absorbing material, and a scattering material. In this paper, we focused on the property of scattering, which only depends on the differences of IoR between the base material and scattering material. We measured the material property of the epoxy and decided to use Stycast1090 as a base material. We employed powder beads as a scattering material. The typical size of the powder beads was controlled to be that of the Mie-scattering range for millimeter wave. In this paper, we compared KEK black with Bock black. Thus, we were able to see the significantly higher performance of KEK Black as an absorber.

In this paper, we characterized the optical performance of absorber at millimeter wave length. The potential of powder beads as void in dielectric material might be appeared at other wavelengths.  At wavelengths large compared to the low-density voids, the material can be considered as an effective homogenous dielectric mixture in the mean field theory limit ~\cite{EMT}. The physical effect described in the strategy to realized low reflectance and high absorption coatings is used in visible or infrared region. The similar technology is already used for the commercial application in~\cite{MJP}. The theory for modeling the performance of black absorber might be used from these fields.


We employed the carbon black as absorbing material. carbon lump black is ~30nm nano-particles of carbon soot. The size of the particles, the dielectric function of the host and conductive loading media, and the wavelength affect the performance of KEK black. In this paper, we mentioned the mass of carbon lamp black for each sample. However, we note that understanding of the formulation needs to compare the volume loading fractions. In this paper, we discussed the phenomenological result based on the commercial lump black. The understanding of formulation for lump black is one of the important issue for future work.

\begin{figure}[ht!]
\centering
\includegraphics[width=\linewidth]{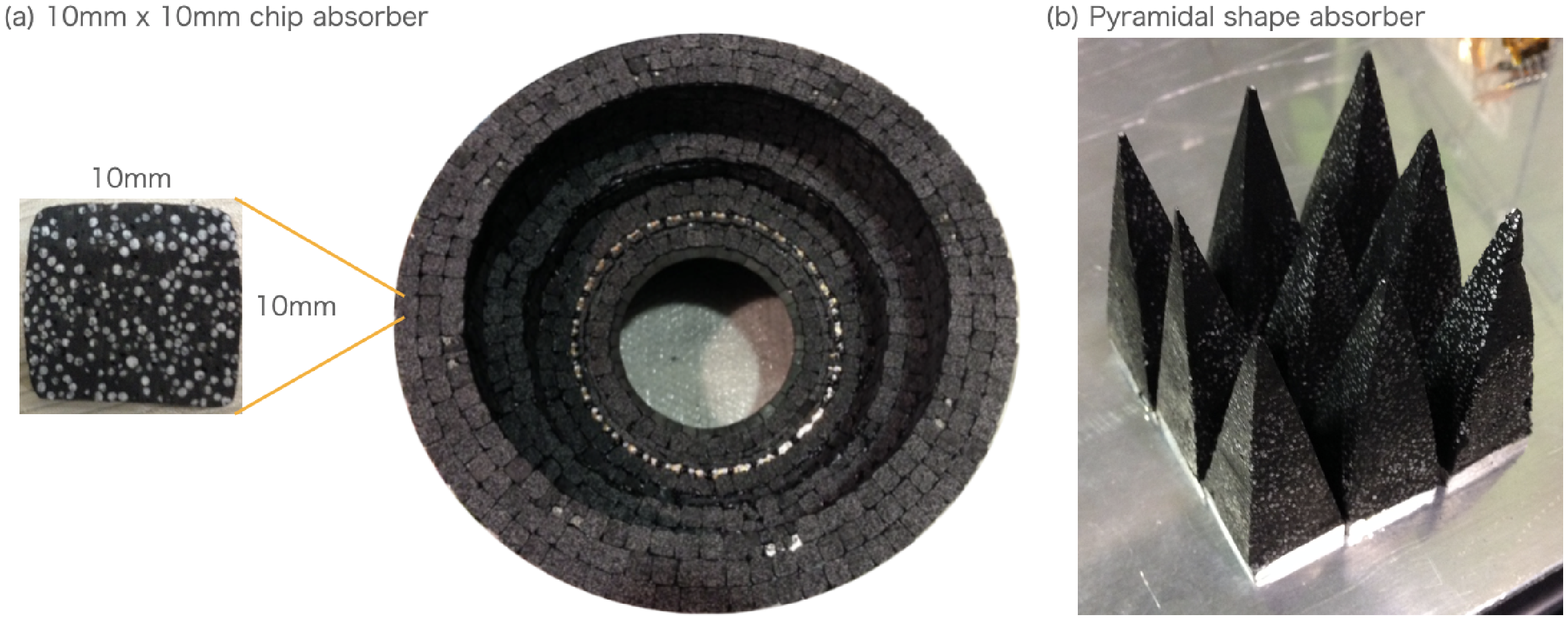}
\caption{The example of application of KEK black. Left figure (a) shows the chip shape absorbers. Right figure (b) shows the pyramidal shape absorber.}
\label{fig11}
\end{figure}

Conductively loaded low-density dielectric open and closed cell foams rely on similar mechanisms for absorption and impedance matching for bulk layers. Low density voids in the material reduce the overall effective dielectric function in the mean field theory limit and the geometry features serve as scatters to improve absorptance as the wavelength approaches. Such media can be stacked or geometrically structured at wavelength scales, such as, through the use of graded index, tapers, cones, wedges, to reduce reflectance observed in a homogenous layer of the media. KEK black has been used in the POLEARBEAR-2 experiment, Simons Array experiment, and Sky Emulator. These experiments applied KEK black under cryogenic temperatures. They cut the KEK black into 10~mm $\times $ 10~mm $\times $ t5~mm chips and put them on a cryogenic surface. The temperature difference between the surface and back surface was less than 1~K, indicating sufficient thermal conductivity for cryogenic application. These experiments also applied KEK black as Lyot stop to define the aperture. In the case of Sky Emulator application~\cite{fixen,ed,6}, they formed a pyramidal shape to increase the performance of multi reflection on the surface.  Figure~\ref{fig11} shows the examples of applications for 10~mm $\times $ 10~mm $\times $ t5~mm chips and pyramidal shape.
The density of KEK black is much lower than bock black, meaning that we could reduce the total mass of the system. A light telescope allows the quick scan to reduce the $1/ f $ noise.
The total mass of the system is more serious for satellite experiments and may be used for LiteBIRD satellite~\cite{20,21,22}. However, we must mention that we need to characterize the performance of radiation resistivity for satellite experiments in future studies.

\section{SUMMARY}
\label{sec5}

We describe and characterize a newly developed absorptive epoxy formulation, `KEK black.' We compare it with conventional ones, where we tried various recipes for producing the KEK black. The performance with any recipe we tried is better than that of the conventional absorber. The reflectance of many powder beads samples tends to be less than that of others. However, the differences of C and D (G and H) are not significant. Therefore, we conclude that sample G has better performance in millimeter wavelengths. The transmittance of sample G is less than 10\% when we use a 2~mm thick sample. The transmittance in sub-millimeter wavelengths is less than 1\%. The averaged reflectance of sample G is less than 20\%, which is sufficiently less than that of the conventional absorber.

\section*{Acknowledgment}

We would like to thank Hiroshi Matsuo and Tom Nitta for their help with THz measurements. We are also grateful to Satoru Igarashi, Iwao Murakami, and Keishi Toyoda for their help in setting up the measurement. We would like to express our gratitude to Adrian Lee, Tomotake Matsumura Akito Kusaka, and Junichi Suzuki. We would like to thank the KEK Cryogenics Science Center and International Center for Quantum-field Measurement System for Studies of the Universe and Particle(QUP)  for its support. Yuki Inoue was supported by Advanced Research Course in SOKENDAI (The Graduate University for Advanced Studies), Academia Sinica, and  the Ministry of Science and Technology (MoST) in Taiwan under Grants No.110-2636-M-008 in Taiwan. KEK authors were supported by MEXT KAKENHI Grant Numbers JP21111002 and JP15H05891 and JSPS KAKENHI Grant Numbers JP13J03626, JP24740182, JP24684017, JP15H03670, JP24111715, and JP26220709. Masashi Hazumi and Masaya Hasegawa were supported by World Premier Intenational Research Center Initiative (WPI), MEXT, Japan.

\section*{Disclosures}
The authors declare no conflicts of interest.

\section*{Data availability}
Data underlying the results presented in this paper are not publicly available at this time but may be obtained from the authors upon reasonable request.


\end{document}